\documentclass[11pt]{article}

\usepackage{amsfonts}
\usepackage{amsmath}
\usepackage{epsf}
\usepackage{graphicx}

\usepackage{xcolor}

\newcommand{\rmi}{\mathrm{i}}
\newcommand{\rmd}{\mathrm{d}}

\newcommand{\bfr}{\mathbf{r}}

\renewcommand{\qquad}{\hspace*{25pt}}

\begin{document}

\begin{center}
{\Large\bf Phases of the Bose-Einstein condensate dark matter model with both two- and three-particle interactions}\\

\vspace*{2.5mm}
{A.M. Gavrilik\footnote{e-mail: omgavr@bitp.kiev.ua} and A.V Nazarenko}\\

\vspace*{1.5mm}
{\small Bogolyubov Institute for Theoretical Physics of NAS of Ukraine\\
14b, Metrolohichna Str., Kyiv 03143, Ukraine}
\end{center}

\abstract{
In this paper we further elaborate on the Bose-Einstein
condensate (BEC) dark matter model extended in our preceding work
[{\em Phys. Rev. D} {\bf 2020}, {\em 102}, 083510] by the inclusion of 6th order (or
three-particle) repulsive self-interaction term. Herein, our goal is
to complete the picture through adding to the model the 4th order
repulsive self-interaction. The results of our analysis confirm the
following: while in the preceding work the two-phase structure and
the possibility of first-order phase transition was established,
here we demonstrate that with the two 
self-interactions involved, the nontrivial phase structure of the enriched
model remains intact. For this to hold, we study the conditions which
the parameters of the model, including the interaction parameters,
should satisfy. As a by-product and in order to provide some
illustration, we obtain the rotation curves and the (bipartite)
entanglement entropy for the case of particular dwarf galaxy.}


\section{Introduction}

Although the concept of dark matter (DM) is a widely accepted one, its
precise nature is still escaping. There exist vast multitude of
different approaches and models, among which the modeling of DM as
Bose-Einstein condensate (BEC), see e.g.~\cite{Sin94,Lee96,HuBG,BH07},    
the overviews~\cite{SRM14,Fan16} and many others, finds each time
more and more support. Positions of BEC model of DM were especially
enforced after the works ~\cite{Harko11,DHNM180,Harko14} which demonstrated ability of BEC
DM to avoid the core-cusp and the gravitational collapse \cite{KMZ} problems.
Also, the model gives quite successful
description~\cite{Harko11,CHarko20,MM12,Harko18} of the rotation curves of
a number of galaxies, at least dwarf and the low surface brightness ones.

Nevertheless, even within this well-elaborated 
model there also exist some tensions and issues which can be improved.
To this end, there is in particular a possibility to apply appropriate tools
from the powerful and efficient theory of deformations. Namely, the $\mu$-deformed
analog of Bose-gas model developed in~\cite{GKkhN18}, with the so-called
$\mu$-calculus as a base, has demonstrated         
clearly the following preferable features: (i)
the evaluated mass of DM halo appears more realistic; (ii) the obtained
critical temperature of condensation of $\mu$-Bose gas $T^{(\mu)}_C$ depends on the deformation
parameter $\mu$, \ $\mu>0$, and is higher~\cite{RGK13} than the usual $T_C$;
(iii) $\mu$-deformation based description of the rotation curves~\cite{GKkh19}
fits better than the curves inferred within the ordinary BEC model.

It is also worth to mention the recent work~\cite{Naz20} which uses the concept of
deformed spatial commutation relations for scalar field in order to develop
a class of generalization of the Bose-condensate DM model. Such an extension has good
potential to achieve improvements.

Another line of extension of the BEC model of DM, developed recently in~\cite{GKN20},
involves sixth order (or 3-particle) self-interaction term ~$\psi^6$.
 Due to presence of the latter, the modified model manifests         
nontrivial phase structure: there exist two distinct phases,
certain region of instability, and the possibility of first-order phase transition.

In order to make the extended model even more complete, it is natural
to include, besides the $\psi^6$, also the two-particle self-interaction
encoded in the term ~$\psi^4$.   Analysis of such ``doubly-nonlinear''
extension of~\cite{GKN20} and BEC model of DM is the goal of the present paper.

The structure of the paper is the following. Necessary
details of the model are given in Section~2, and main part that involves
obtaining thermodynamic functions and their key properties is
presented in Section~3. In Sections 4 and 5 we consider briefly, again
for the situation of presence of both two- and three-particle
interactions, the rotation curves of selected galaxy and the
respective bipartite entanglement entropy of two centrally symmetric
regions of the halo of this same galaxy.
 In our concluding section, we present a discussion of the results.

\section{The Model}

We describe the Bose-Einstein condensate by real function $\psi(r)$ of radial variable $r=|\bfr|$,
using a constant chemical potential $\tilde\mu$. Our study is based on the energy functional $\Gamma$
in a ball $B=\{\bfr\in\mathbb{R}^3|\, |\bfr |\leq R\}$ and the Poisson equation:
\begin{eqnarray}
&&\hspace*{-9mm}
\Gamma=4\pi\int_0^R\left[\frac{\hbar^2}{2m}(\partial_r\psi(r))^2
+m\psi^2(r)V_{\mathrm{gr}}(r)+\frac{U_2}{2}\psi^4(r)
+\frac{U_3}{3}\psi^6(r)-\tilde\mu\psi^2(r)\right]\,r^2\,\rmd r,\label{G1}\\
&&\hspace*{-9mm}
\Delta_r V_{\mathrm{gr}}(r)=4\pi Gm|\psi(r)|^2,
\end{eqnarray}
where $\Delta_r$ is the radial part of Laplace operator that acts as
\begin{eqnarray}
&&\Delta_rf(r)=\partial^2_rf(r)+\frac{2}{r}\,\partial_rf(r),\label{Dlt}\\
&&\Delta^{-1}_rf(r)=-\frac{1}{r}\int_0^rf(s)\,s^2\,\rmd s-\int_r^{R}f(s)\,s\,\rmd s,
\label{InvDelta}
\end{eqnarray}
$R$ being the radius of the ball where the matter is located.

Focusing here on the effects of interparticle interactions,
we leave aside the slow rotation of the
condensate~\cite{Harko18}, which can be taken into account 
through the chemical potential~\cite{Naz20}.

For convenience, let us introduce dimensionless variables:
\begin{eqnarray}
&&\psi(r)=\sqrt{\varrho_0}\,\chi(\xi),\quad r=r_0\,\xi,\quad u=\tilde\mu\,\frac{m r_0^2}{\hbar^2},\nonumber\\
&&A=4\pi\frac{Gm^3\varrho_0 r_0^4}{\hbar^2},\quad
Q=U_2\frac{\varrho_0 m r_0^2}{\hbar^2},\quad
B=U_3\frac{\varrho_0^2 m r_0^2}{\hbar^2}.
\label{param}
\end{eqnarray}
Here $\chi(\xi)$ is a real dimensionless scalar
field; $\varrho_0$ and $r_0$ characterize {\it typical measures} of
the central particle density and the system size, respectively.

Thus, we arrive at
\begin{eqnarray}
&&\hspace*{-5mm}
\frac{\Gamma}{\Gamma_0}=\int_0^{\xi_B}\left[\frac{1}{2}(\partial_\xi\chi)^2-u\chi^2
+A\chi^2\varphi+\frac{Q}{2}\chi^4+\frac{B}{3}\chi^6\right]\xi^2\,\rmd\xi,\quad
\Gamma_0=\frac{4\pi\hbar^2r_0\varrho_0}{m},
\nonumber\\
&&\hspace*{-5mm}
\Delta_{\xi}\varphi(\xi)=\chi^2(\xi),\label{G2}
\end{eqnarray}
where $\xi_B=R/r_0$, whereas $\Delta_{\xi}$ and $\Delta^{-1}_{\xi}$ are given by (\ref{Dlt}),
(\ref{InvDelta}) in terms of $\xi$ replacing $r$.

To estimate the range of parameter values, we turn to astrophysical situations.
Since we suggest to take into account the three-particle interaction in relatively
dense DM of light bosons with masses of the order of $10^{-22}\ \mathrm{eV}\,c^{-2}$,
ranges of the parameters can be found by considering the DM of galactic cores
with a central mass density $\rho_0=m\varrho_0$ of the order of $10^{-20}\ \mathrm{kg}\,\mathrm{m}^{-3}$,
in the region of radius $r_0$ smaller than 1~kpc. Then, extracting $r_0$ from the definition (\ref{param})
of the measure of gravitational interaction $A$,
\begin{equation}
r_0\simeq 0.824\ \mathrm{kpc}\,\left[\frac{A}{10}\right]^{1/4}
\,\left[\frac{mc^2}{10^{-22}\ \mathrm{eV}}\right]^{-1/2}\,
\left[\frac{\rho_0}{10^{-20}\ \mathrm{kg}\,\mathrm{m}^{-3}}\right]^{-1/4},
\label{rA}
\end{equation}
we can adopt that $A\sim10$ \cite{GKN20}.

It is clear that the gravity results from integral effect of a whole system.
Unlike, (thermo)dynamics of internal processes is determined by repulsive interactions among bosons,
represented by the parameters $Q$ and $B$ under the condition $B>A$. The role of pairwise
interaction, controlled by $Q$, is assumed to be comparable with the effect of gravity.

Using (\ref{rA}), the characteristic energy density $\varepsilon_0=\hbar^2\varrho_0/(m r^2_0)$
is evaluated as
\begin{equation}
\varepsilon_0\simeq 33.82\ \mathrm{eV}\,\mathrm{cm}^{-3}\,\left[\frac{A}{10}\right]^{-1/2}
\,\left[\frac{mc^2}{10^{-22}\ \mathrm{eV}}\right]^{-1}\,\left[\frac{\rho_0}{10^{-20}\ \mathrm{kg}\,\mathrm{m}^{-3}}\right]^{3/2}.
\label{eA}
\end{equation}
In the pressure units, $33.82\ \mathrm{eV}\,\mathrm{cm}^{-3}\simeq 5.42\cdot10^{-12}\ \mathrm{Pa}$.

Extremizing of functional $\Gamma$, i.e. $\delta\Gamma/\delta\chi(\xi)=0$, yields the set of field equations:

\begin{equation}\label{teq1}
\frac{1}{2}\Delta_\xi\chi+u\chi-A\chi\varphi-Q\chi^3-B\chi^5=0,\qquad
\Delta_{\xi}\varphi=\chi^2.
\end{equation}
We combine the model equations in the spirit of \cite{GKN20} by
introducing the field $\upsilon(\xi)$:
\begin{equation}
\upsilon(\xi)=\int_0^{\xi}\chi^2(s)\,s\,\rmd s,\qquad
\upsilon(\xi_B)=-\varphi(0).
\end{equation}
As result,
\begin{eqnarray}
&& 2\frac{\Gamma}{\Gamma_0}=\int_0^{\xi_B}\left[(\partial_\xi\chi)^2-u_*\,\chi^2(\xi)
+\frac{Q_*}{2}\,\chi^4(\xi)+\frac{B_*}{3}\,\chi^6(\xi)\right]\,\xi^2\,\rmd\xi\nonumber\\
&&\hspace*{10mm}-\frac{A_*}{2}\int_0^{\xi_B} [\upsilon(\xi_B)-\upsilon(\xi)]^2\,\rmd\xi,\label{g1}\\
&& \Delta_\xi\chi+\nu\chi-\chi \frac{A_*}{\xi}\int_0^\xi\upsilon(s)\,\rmd s-Q_*\chi^3-B_*\chi^5=0,\label{g2}\\
&& \partial_\xi\upsilon(\xi)=\xi\,\chi^2(\xi),\qquad \upsilon(0)=0,\label{g3}\\
&& \nu=A_*\upsilon(\xi_B)+u_*,\label{g4}
\end{eqnarray}
where $A_*=2A$, $Q_*=2Q$, $B_*=2B$, and $\nu$ (put instead of $u_*=2u$) are
arbitrary positive parameters. The system boundary $\xi_B$ is
defined from the condition $\chi(\xi_B)=0$ and is the {\it first zero}
of oscillating function $\chi(\xi)$.

In order to find a decreasing solution $\chi(\xi)$ for admissible $\xi$ with a finite initial value
$\chi_0=\chi(0)<\infty$, we first impose $\chi^\prime(0)=0$ and then formulate the conditions which allow to fix $\chi_0$,
through expanding $\chi(\xi)=\chi_0+C_2\xi^2+\dots$ at $\xi\to0$. On substituting that in (\ref{g2}), (\ref{g3}),
the following algebraic equations result:
\begin{eqnarray}
&&6C_2+\nu \chi_0-Q_* \chi^3-B_* \chi^5_0=0,\nonumber\\
&&\nu C_2-\frac{A_*}{6}\,\chi^3_0-3Q_* \chi^2_0 C_2-5B_* \chi^4_0 C_2=0.
\label{ICs}
\end{eqnarray}
Combining the two, we find that the value $\chi_0$ should satisfy
the equation $S(A_*,B_*,Q_*,\nu,\chi_0)=0$, where

\begin{equation}
S(A_*,B_*,Q_*,\nu,z)=A_* z^2-(5B_*z^4+3Q_*z^2-\nu)(\nu-Q_*z^2-B_*z^4),
\end{equation}
plus the condition $2C_2=\chi^{\prime\prime}(0)\leq0$.
Taken altogether, these constrain $\chi_0$ as $z_1<\chi_0<z_2$, where
\begin{equation}
z_1=\left[\sqrt{\left(\frac{3Q_*}{10B_*}\right)^2+\frac{\nu}{5B_*}}-\frac{3Q_*}{10B_*}\right]^{1/2},\quad
z_2=\left[\sqrt{\left(\frac{Q_*}{2B_*}\right)^2+\frac{\nu}{B_*}}-\frac{Q_*}{2B_*}\right]^{1/2}.
\end{equation}
Technically, the search for the initial value $\chi_0$ of the model
which takes into account the pair interaction is similar to the problem with
three-particle interaction only~\cite{GKN20}. Likewise, we notice three regimes (for given
$A_*$, $Q_*$, $B_*$ and $\nu$): 1) no solution for $\chi_0$ that, in Eq.~(\ref{g2}), leads to $\chi(\xi)=0$;
2) single solution $\chi_0$ that corresponds to a {\it minimal admissible value} $\nu_{\mathrm{min}}$ from which
the system starts to evolve; 3) pair of (positive) solutions for $\chi_0$, when we should choose
a minimal one, because the other leads to divergent $\chi(\xi)$. Usually, for fixed $(A_*,Q_*,B_*)$,
but increasing $\nu$, the indicated sequence of all three options occurs.

\begin{figure}
\begin{center}
\includegraphics[width=6.5cm]{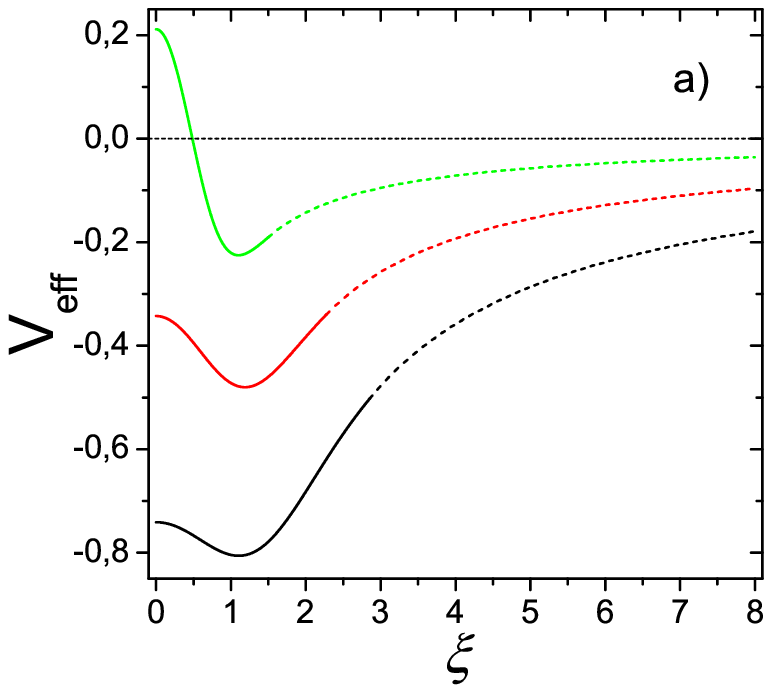}\ \ \
\includegraphics[width=6.5cm]{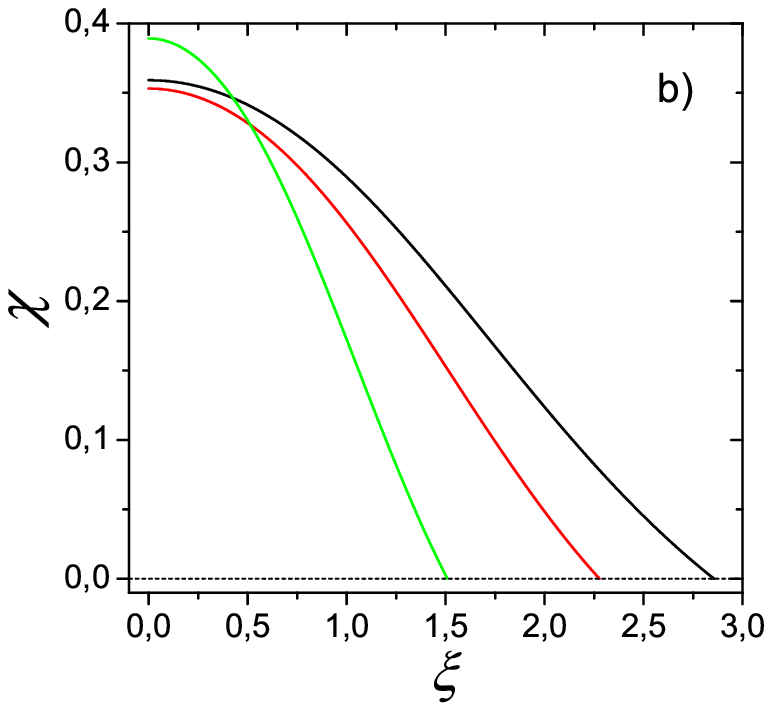}
\end{center}
\caption{\label{V}\small a) The effective potential $V_{\mathrm{eff}}(\xi)$ for $u_{\mathrm{black}}\simeq0.149$,
$u_{\mathrm{red}}\simeq-0.502$, $u_{\mathrm{green}}\simeq-2.046$. Dashed parts of the curves correspond to a pure gravitational
potential $-A\cdot {\cal N}/\xi$ outside the particle system. b) The field $\chi(\xi)$ for the same
values of chemical potential $u$, and $\chi(\xi)=0$ at $\xi=\xi_B$. Here $A=10$, $B=20$, $Q=1.36$
for definiteness.}
\end{figure}

It is useful to analyze the system from the quantum-mechanical point of view.
Equation (\ref{g2}) for $\xi\leq\xi_B$ can be conveniently rewritten in the Schr\"odinger
form to describe the scattering of a particle whose wave function\footnote{See Fig.~\ref{V}b
for the behavior of $\chi(\xi)$.} is taken as $f(\xi)=c\chi(\xi)$ ($c$ is a normalization depended
on a total number of particles) in the potential $V_{\mathrm{eff}}$:
\begin{eqnarray}
&&\left(-\frac{1}{2}\Delta_\xi+V_{\mathrm{eff}}(\xi)\right) f(\xi)=u f(\xi),\label{Sch1}\\
&&V_{\mathrm{eff}}(\xi)=V_2(\xi)+V_3(\xi),\\
&&V_2(\xi)=Q\chi^2(\xi)+V_{\mathrm{gr}}(\xi), \quad V_3(\xi)=B\chi^4(\xi),\\
&&V_{\mathrm{gr}}(\xi)=-A\upsilon(\xi_B)+\frac{A}{\xi}\int_0^\xi\upsilon(s)\rmd s,\label{eV4}
\end{eqnarray}
where potentials $V_2$ and $V_3$ come from two and three-particle interactions, respectively.
Substituting the found solution $\chi(\xi)$, we can see that the form of potential $V_{\mathrm{eff}}$
depends also on a chemical potential $u$ (or parameter $\nu$).

In Fig.~1 we have used the values $A=10$, $B=20$ and $Q=1.36$. The latter one plays the role
of  ``critical'' value (its sense is seen in Fig.~2 below, with explanations at the end of the
next Section). We relate the particular forms of $V_{\mathrm{eff}}$ given in Fig.~\ref{V}a with
the physical situations depicted in Fig.~\ref{PT} below.
Namely, {\it green curve} in Fig.~\ref{V}a is chosen for liquid-like (dense) state in Fig.~\ref{PT},
when the three-particle interaction contributes to a hard-core part of potential at small $\xi$.
{\it Red curve} is constructed in the vicinity of the critical point of the first-order phase transition,
when the potential $V_{\mathrm{eff}}$ is similar to the harmonic trap. {\it Black curve}
corresponds to gaseous (dilute) state in the effective (truncated) gravitational potential,
when the kinetic energy term dominates ($u>0$).

In the range $\xi\geq\xi_B$ we come to the problem of a particle in the gravitational field
(see the dashed curves in Fig.~\ref{V}a) created by the system of ${\cal N}$ particles:
\begin{equation}
\left(-\frac{1}{2}\Delta_\xi-A\frac{{\cal N}}{\xi}\right) f_k(\xi)=\frac{k^2}{2} f_k(\xi),\quad
f_k(\xi_B)=0,\quad f^\prime_k(\xi_B)=f^\prime(\xi_B),
\end{equation}
where
\begin{equation}
{\cal N}=\int_0^{\xi_B}\chi^2(\xi)\,\xi^2\rmd\xi
\end{equation}
is the total number of particles within the ball $\xi\leq\xi_B$, which determines the total mass.

At this stage a wave number $k$ is ambiguous. Oscillating and decaying
solution to this equation (for any real $k$ and pure imaginary $\kappa$ and $s$) is given as:
\begin{eqnarray}
&&f_k(\xi)=\frac{c_1 M_{s,1/2}(\kappa\xi)+c_2 W_{s,1/2}(\kappa\xi)}{\xi},\quad
\kappa=2\rmi k,\quad s=\frac{2A{\cal N}}{\kappa},\\
&&c_1=f^\prime(\xi_B)\,\xi^2_B\, \frac{W_{s,1/2}(\kappa\xi_B)}{F(\kappa)},\qquad
c_2=-f^\prime(\xi_B)\,\xi^2_B\, \frac{M_{s,1/2}(\kappa\xi_B)}{F(\kappa)},\\
&&F(\kappa)=(1+s)M_{1+s,1/2}(\kappa\xi_B)\,W_{s,1/2}(\kappa\xi_B)
+M_{s,1/2}(\kappa\xi_B)\,W_{1+s,1/2}(\kappa\xi_B).
\end{eqnarray}
Here $M_{\mu,\nu}(z)$ and $W_{\mu,\nu}(z)$ are the Whittaker functions.

This solution might be interpreted as the gravitational capture of a dark matter particle
(a radial wave with energy $k^2/2$) outside the dark matter ball, when $\xi\to\infty$.
In the case of initially resting particle with $k=0$, the solution is described in terms of
Bessel functions $J_1(z)$ and $Y_1(z)$.

Thus, the account of interaction confirms the possibility of bound states,
which can manifest themselves in the form of different thermodynamic phases. Further,
all global quantities of the model, computed at fixed $A$, $Q$ and $B$, are supposed
to be functions of free parameter $\nu$. Therefore, dependence, say, of $a$ on $b$ should be
treated in parametric form: $a(b)=\{(b(\nu),a(\nu))|\nu\geq\nu_{\mathrm{min}}\}$.

\section{Thermodynamic Quantities and Two Phases}

To study the macroscopic properties, let us define an effective chemical potential $\mu(\xi)$ \cite{LL},
which includes the gravitational potential jointly with the term of quantum fluctuations, and replaces further
the constant chemical potential $u$, that is
\begin{equation}
\mu(\xi)+A\varphi(\xi)-\frac{1}{2\chi(\xi)}\Delta_\xi\chi(\xi)=u.
\end{equation}
Accordingly to the equation of motion (\ref{teq1}), $\mu$ determines
$\chi$ as
\begin{equation}\label{muQ}
\mu(\xi)=Q\chi^2(\xi)+B\chi^4(\xi),\qquad \mu(\xi_B)=0.
\end{equation}

For finding macroscopic characteristics, we appeal to the thermodynamic relations at $T=0$,
using a local particle density $\eta(\xi)=\chi^2(\xi)$:
\begin{eqnarray}
\rmd p(\xi)&=&\eta(\xi)\,\rmd\mu(\xi),\hspace*{17.2mm} p(\xi_B)=0,\label{GDr}\\
\varepsilon(\xi)&=&\eta(\xi)\,\mu(\xi)-p(\xi),\qquad \varepsilon(\xi_B)=0,\label{Eu}
\end{eqnarray}
where functions $p(\xi)$ and $\varepsilon(\xi)$ determine the (dimensionless)
mean pressure $P$ and the internal energy $E$:

\begin{equation}
P=\frac{3}{\xi^3_B}\int_0^{\xi_B}p(\xi)\,\xi^2\,\rmd\xi,\qquad
E=\int_0^{\xi_B}\varepsilon(\xi)\,\xi^2\,\rmd\xi.
\end{equation}
Hereafter, $\xi^3_B/3$ represents the volume of the system.

Therefore, we need to integrate the Gibbs--Duhem relation (\ref{GDr}) and then
to substitute $p(\xi)$ into the Euler relation (\ref{Eu}) in order to find $\varepsilon(\xi)$.
This way, the explicit expressions are obtained:
\begin{equation}\label{EPtf}
p(\xi)=\frac{Q}{2}\eta^2(\xi)+\frac{2}{3}B\eta^3(\xi),\qquad
\varepsilon(\xi)=\frac{Q}{2}\eta^2(\xi)+\frac{1}{3}B\eta^3(\xi),
\end{equation}
which give us the equation of state by inserting the solution $\eta(\xi)$ of (\ref{g2})--(\ref{g4}).

The (local) internal pressure $p(\xi)$ behaves spatially with radius accoding to
\begin{equation}\label{qPev}
\frac{\partial_\xi p(\xi)}{\eta(\xi)}=\partial_\xi\mu_q(\xi)=-A\frac{n(\xi)}{\xi^2}
+\partial_\xi\left(\frac{1}{2\chi(\xi)}\Delta_\xi\chi(\xi)\right),\quad
n(\xi)=\int_0^\xi\eta(s)\,s^2\,\rmd s.
\end{equation}
We do not solve it because all necessary fields are found explicitly from
Eqs.~(\ref{g2})--(\ref{g4}).

Mean particle density is
\begin{equation}
\sigma=\frac{3}{\xi^3_B}\int_0^{\xi_B}\chi^2(\xi)\,\xi^2\rmd\xi.
\end{equation}

\begin{figure}
\begin{center}
\includegraphics[width=8cm]{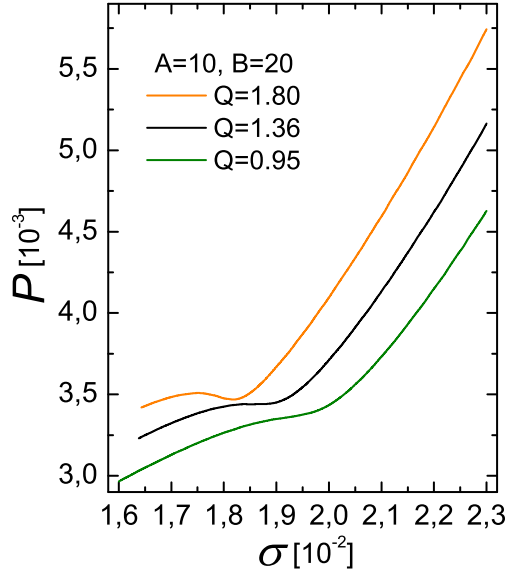}
\end{center}
\caption{\label{PT}\small Equation of state of dark matter at $T=0$ and fixed parameters $A$ and $B$.
A slow growth in dimensionless pressure $P$ at relatively low density $\sigma$ corresponds to a dilute
phase of matter, while a steep rise indicates a denser liquidlike phase at high densities. Orange curve
(with two extrema) demonstrates
the presence of metastable states. Black line is for a critical value of $Q_c=1.36$. Green line
exemplifies a continuous transition between the two phases at $Q<Q_c$.}
\end{figure}   

Dependence of mean internal pressure $P$, see (31), on this density $\sigma$
is presented in Fig.~\ref{PT} for various parameters of interaction. The curves obtained numerically
reveal the presence of two stable phases of dark matter with $\partial P/\partial\sigma>0$:
the dilute one ($\partial^2P/\partial\sigma^2<0$) and the denser liquidlike one
($\partial^2P/\partial\sigma^2>0$).
The characteristic points of the phase diagram are determined
from the conditions: $\partial P/\partial\sigma=0$ and $\partial^2P/\partial\sigma^2=0$.
The simultaneous fulfillment of these conditions at a single point determines the critical point,
which belongs to the black curve in Fig.~\ref{PT}, and inferring of which is one of main tasks here.
The existence of a critical point of a first-order phase transition is essentially related
to the competition between gravitational and pair interactions, controlled by the parameter $Q$.

While the presence of metastable states, linked with the two-extrema behavior at $Q>1.36$
and shown by the orange curve, clearly indicates mixing of the gaseous
and liquid phases, a detailed description of the transition between the two phases at $Q<1.36$ requires
the use of additional thermodynamic characteristics, as is already done in \cite{GKN20}
by means of the {\em perturbation pressure} $\Pi_\nu$ for $Q=0$.

It is important to emphasize that herein we reveal a discontinuous behavior
(jump between two phases) of the density $\sigma$ with the change in the {\em internal pressure} $P$.
The existence of such a regime, as we show, is allowed at $Q>1.36$ for $A=10$ and $B=20$.
Although the existence of a denser, liquidlike phase of DM (in a relatively small region of halo)
is possible at $Q=0$, the condition $Q>0$ is required in describing the galactic DM
halos~\cite{Harko11}, even without taking into account the three-particle interaction.

Note that a continuous change in the parameter $Q$ can lead to the intersection
of different curves $P(\sigma)$, that indicates the possibility of realizing one thermodynamic
state by fixation of different sets of parameters.
Such ambiguity in the set of parameters can complicate the interpretation and reproduction of
observables.

\section{The Rotation Curves}

Important information about dark matter is extracted and verified
from the rotation curves of galaxies.
As our model is aimed
to study the processes in dark matter in the galaxy cores, that is,
in relatively small regions of space with a noticeable density, its direct
application to the description of rotation curves should be limited
to dwarf galaxies (or other dark matter dominated compact galaxies).
One of the possibilities for describing larger (halo-type)
objects is to extend the model, similarly to what was proposed in \cite{GKN20}.
Besides, the lack of accounting for the rigid
rotation of matter in the model does not allow us to describe the
curves of rotating galaxies. Nevertheless, we consider it important
and interesting to demonstrate the capabilities of our model, in
which we accurately took into account quantum fluctuations in the
condensate and (both two- and) three-particle
interactions. Although the extensions of the model are indicated in
\cite{GKN20,Naz20}, the included effects make it possible to
further validate the Bose-condensate approach, to
compare with and supplement the results of \cite{Harko11}.

Thus, the tangential velocity $v$ of a test particle moving in the spherically
symmetric DM halo can be represented as

\begin{equation}
v(r)=\sqrt{G\frac{M(r)}{r}},\qquad
M(r)=4\pi\int_0^r\rho(s)\,s^2\,\rmd s,
\end{equation}
where $\rho(r)=m|\psi(r)|^2$ is a mass density such that $\rho(R)=0$ at $R=r_0\xi_B$.

\begin{figure}
\begin{center}
\includegraphics[width=11cm]{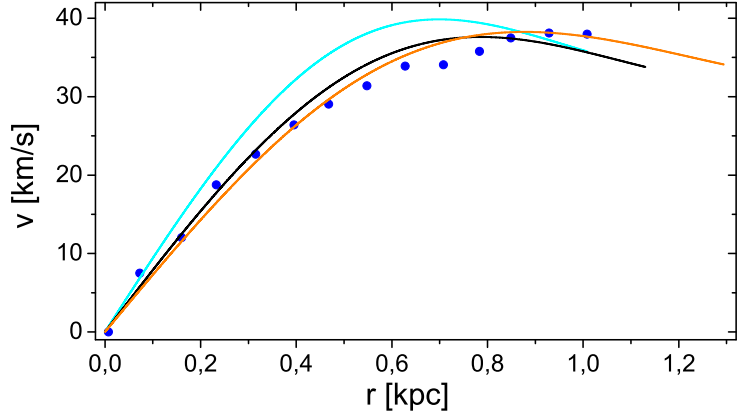}
\end{center}
\caption{\label{M81}\small Rotation curves for dwarf galaxy M81dwB with mass $M=3\cdot10^8 M_\odot$.
{\it Blue dots} are the observed data for $r\leq1~\text{kpc}$. Solid lines are obtained within
the model under certain restrictions: {\it cyan line} is for $M^{(1)}_{\rm BEC}=3\cdot10^8 M_\odot$
 and $R^{(1)}=1~\text{kpc}$; {\it black line} is for $M^{(2)}_{\rm BEC}=3\cdot10^8 M_\odot$
 and $R^{(2)}\simeq1.12~\text{kpc}$; {\it orange line} corresponds to $M^{(3)}_{\rm BEC}=3.5\cdot10^8 M_\odot$
 and $R^{(3)}=1.3~\text{kpc}$.}
\end{figure} 

Let us illustrate the model predictions for the rotation curves of the M81 galaxy,
shown in Fig.~\ref{M81}.
Note especially the free parameters of the model $A$, $B$, $Q$, $\nu$, $\rho_0$ that we use for fitting,
as well as the restricting characteristics: the total mass $M$ of dark matter and the halo radius $R$.

The galaxy M81 with no rigid rotation gives us the most striking
proof of the Bose-con\-den\-sa\-te approach, as noted in
\cite{Harko11}. Let us use this example to   
emphasize main features of the description of rotation
curves.

First of all, note it looks rather difficult to
indicate unambiguously the  
parameters for the rotation curve: the same dependence $v(r)$ can be
realized for different sets of the parameters.
This is a consequence of the symmetry of the model equations.
For this reason, we give only graphs that are interesting for physics,
and omit the mathematics of identifying the symmetries.

In our example, it is worth to note a possibility to
construct the curve colored in cyan in Fig.~\ref{M81} with $Q=0$, when
the pair interaction is absent. At the same time, the best fitted
dependencies, represented by other curves, always require $Q>0$.
This implies that the repulsive pair interaction should be taken into
account in realistic models of dark matter in (dwarf) galaxies~\cite{CHarko20,KKG}.

\section{Entanglement Entropy}

The possibility of estimating the entanglement entropy in a
general Bose condensate was developed, in particular,
in \cite{KRS06}. The idea of using this entropy as a criterion for
differentiating the BEC dark matter from
cold dark matter (CDM) was proposed in \cite{Lee18}.
 Although a more precise analysis requires studying the effects
 of interference of particles from
interacting subsystems, here we estimate the entropy between two
separated (central and surrounding) parts of radially inhomogeneous
BEC dark matter of a galaxy. Omitting the normalization
constants, we use the formula (in accordance with
\cite{KRS06,Lee18}):
\begin{equation}
S_E(x)=\ln{\{c(x\xi_B)\,[1-c(x\xi_B)]\}}, \qquad x=\xi/\xi_B=r/R.
\end{equation}
Here $c(\xi)=n(\xi)/{\cal N}$ is the fraction of particles
contained in the central subsystem, $n(\xi)$ being defined in (33).

\begin{figure}
\begin{center}
\includegraphics[width=8cm]{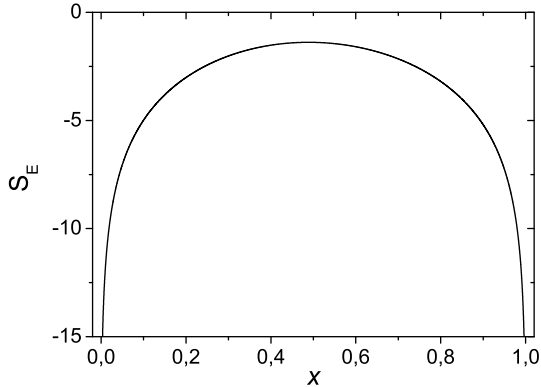}
\end{center}
\caption{\label{En}\small Entanglement entropy $S_E$ within the model with
$A=10$, $B=20$, $Q=1.36$ and $u\simeq-2.046$.}
\end{figure}

A typical dependence of the entanglement entropy on the specific radius is shown
in Fig.~\ref{En}. It is worth noting that, although
the system is inhomogeneous, and $c(\xi)$ is not a
constant, the maximum influence of one subsystem of particles upon the other 
turns out to occur at $x=0.5$. What concerns negative sign of $S_E(x)$:
that is eliminated by explicit account of normalization constant, that is,
by adding definite positive number (of the order $10^2$ or higher, see e.g.
\cite{Lee18}) which depends on the mass of dark matter particles and
their number in the subregion.

\section{Discussion}

In the framework of the extended version of BEC dark matter model
that involves two- and three-particle interactions we explored the
solutions of the system of equations (\ref{g1})-(\ref{g4})
as well as the properties of thermodynamic functions. The analysis
has led us to the conclusions that the interplay of the basic free
parameters $A$, $Q$ and $B$ gives interesting consequences for the
system under study including dark matter halo properties:
 (i) first, the effective interaction potential, see (\ref{Sch1})-(\ref{eV4})
 and Fig.~1, and the density profiles; (ii) the equation of
 state which shows basic dependence on the parameter $Q$
 including existence of its ``critical'' value $Q_C$ that
 separates different regimes; (iii) persistence at $Q>Q_C$ of
 nontrivial phase structure (inherited from the restricted
 pure $\psi^6$ sub-model).  The obtained information on the
 thermodynamic function such as the mass density profile,
 enabled us to infer in Sec.~3 the galactic rotation curves, and
 the bipartite entanglement entropy in Sec.~4. The former clearly
 demonstrated, with the example of dwarf galaxy M81, that the
 model can easily provide nice agreement with observational data.
As follows from the treatment of rotational curves, the parameter
 $Q$ responsible for the two-particle interaction plays essential
 role. Besides, as seen in Fig.~\ref{PT}, at $Q>Q_C$ the model reveals the most rich behavior:
the curves possess two extrema that implies the metastability region.
At $Q\geq Q_C$, only inflection points do survive. Without $Q$ 
(without pair interaction) we could not have such features of
unexpected behavior. Anyway, for both $Q>0$ and $Q=0$ (that brings us back
to the situation explored in \cite{GKN20}) the two phases are present,
though their identification involves differing thermodynamical functions.


It is of interest to compare the model considered in this paper with
some of the nonlinear (namely, non-polynomial) extensions of the BEC
model of DM, e.g. those studied in~\cite{SW,MU,Chav,Chav2} where the
scalar potential $V_0 [\cosh\left(\lambda\kappa\Phi\right)-1]$ was
employed, with $\kappa=\sqrt{8\pi G}$. The essential feature of
these models is the presence of all-order nonlinearities, with the
corresponding powers of the single free parameter $\lambda$ (note
that $\lambda$ can be viewed as a deformation parameter since at
$\lambda \to 0$ the potential and thus the interaction are
vanishing). On one hand, the model studied above is obviously
simpler than the models involving $\cosh$ or $\cos$: indeed, we
encounter the first terms of series expansion if the parameters are
restricted as $A=\sqrt{B}=\lambda$. On the other hand, the
$\sim\!\psi^4$ plus $\sim\!\psi^6$ model studied herein is richer in
the sense that it operates with the free parameters $A,\ {B}, \ Q$,
instead of single one. It is this property that allowed us to
disclose (confirm) the existence of two phases and of the phase
transition. Moreover, the influence of different values of these
parameters was essential in our treatment of the rotation curves and
the entanglement entropy, see Fig.~\ref{En} above.

Concerning the entanglement, an interesting question arises for the
situation when dark matter particles are not elementary bosons, but
composites built from two bosons or two fermions. In the both cases
(i) the composites differ from pure bosons and are naturally
realizable, as can be seen in~\cite{GKM11}, through deformed
oscillators or deformed bosons (or quasi-bosons); (ii) bipartite
 internal entanglement entropy of quasibosons obtained in~\cite{GM12,GM13}
 turned out to depend on the deformation parameter. The question now
 is as follows: to which extent the microscopic intra-quasibosonic
 entanglement and its entropy do affect (superimpose with) the
 macroscopic entanglement and the entanglement entropy that was
 studied in Sec.~5 above. Also it is no doubt important to
 explore in detail the connection~\cite{KRS06,GLRK,WSL,WSLS} between
 peculiarities of the behavior of entanglement and the phase
 transition (of the first order in our case), especially in the
context of the properties of dark matter.  We hope to explore these
questions in one of our future works.

\vspace{6pt} 

\section*{Acknowledgments}

{A.M.G. acknowledges support from the National Academy of Sciences of
Ukraine by its priority project No. 0120U100935 ”Fundamental
properties of the matter in the relativistic collisions of nuclei
and in the early Universe”. The work of A.V.N. was supported by the
project No. 0117U000238 of NAS of Ukraine.}


\end{document}